\documentstyle[11pt]{article}
\newlength{\bredde}
\def\slash#1{\settowidth{\bredde}{$#1$}\ifmmode\,\raisebox{.15ex}{/}
\hspace*{-\bredde} #1\else$\,\raisebox{.15ex}{/}\hspace*{-\bredde} #1$\fi}
\newcommand{\beq}{\begin{equation}}
\newcommand{\eeq}{\end{equation}}
\newcommand{\noi}{\vspace{12pt}\noindent}
\newcommand{\lG}{\raise.3ex\hbox{$\stackrel{\leftarrow}{G}$}}
\newcommand{\lU}{\raise.3ex\hbox{$\stackrel{\leftarrow}{U}$}}
\newcommand{\lP}{\raise.3ex\hbox{$\stackrel{\leftarrow}{{\cal P}}$}}
\newcommand{\leta}{\raise.3ex\hbox{$\stackrel{\leftarrow}{\eta}$}}
\newcommand{\lOmega}{\raise.3ex\hbox{$\stackrel{\leftarrow}{\Omega}$}}
\newcommand{\ldr}{\raise.3ex\hbox{$\stackrel{\leftarrow}{\delta^r}$}}
\def\beqn{\begin{eqnarray}}
\def\eeqn{\end{eqnarray}}
\def\sepand{\rule{14cm}{0pt}\and}
\def\gtwid{\raise.3ex\hbox{$>$\kern-.75em\lower1ex\hbox{$\sim$}}}
\def\ltwid{\raise.3ex\hbox{$<$\kern-.75em\lower1ex\hbox{$\sim$}}}

\begin{document}
\topmargin -1.4cm
\oddsidemargin -0.8cm
\evensidemargin -0.8cm
\title{\Large{
Non-Abelian Antibrackets}}

\vspace{0.5cm}

\author{{\sc Jorge Alfaro} \\
Fac. de Fisica \\ Universidad Catolica de Chile\\
Casilla 306, Santiago 22, Chile \\
\sepand
{\sc Poul H. Damgaard}\\
The Niels Bohr Institute\\ Blegdamsvej 17\\ DK-2100 Copenhagen\\
Denmark}
\maketitle
\vfill
\begin{abstract} The $\Delta$-operator of the Batalin-Vilkovisky
formalism is the Hamiltonian BRST charge of Abelian shift transformations
in the ghost momentum representation. We generalize this $\Delta$-operator,
and its associated hierarchy of antibrackets, to that of an arbitrary
non-Abelian and possibly open algebra of any rank. We comment on the
possible application of this formalism to closed string field theory.
\end{abstract}
\vfill
\begin{flushleft}
NBI-HE-95-39 \\
hep-th/9511066
\end{flushleft}
\newpage


\noindent
In order to see how the conventional antibracket formalism of Batalin and
Vilkovisky \cite{BV} can be generalized, it is important to have a
fundamental principle from which this formalism can be derived. As has
been discussed in a series of papers \cite{us,us1}\footnote{The case
of extended BRST symmetry is derived in ref. \cite{sp2}.}, this
principle is that Schwinger-Dyson BRST symmetry \cite{us0} must be
imposed on the full path integral.

\noi
Schwinger-Dyson BRST symmetry can be derived from the local symmetries
of the given path integral measure. When the measure is flat, the relevant
symmetry is that of local shifts, and the resulting Schwinger-Dyson BRST
symmetry leads directly to a quantum Master Equation on the action $S$
which is exponentiated inside the path integral. This action depends
on two new sets of ghosts and antighosts, $c^A$ and $\phi^*_A$ \cite{us}.
The conventional Batalin-Vilkovisky formalism for an action $S^{BV}$
follows if one substitutes $S[\phi,\phi^*,c] = S^{BV}[\phi,\phi^*]
-\phi^*_Ac^A$ and integrates out the ghosts $c^A$.
The so-called ``antifields'' of the Batalin-Vilkovisky formalism are
simply the Schwinger-Dyson BRST antighosts $\phi^*_A$ \cite{us}.

\noi
It is of interest to see what happens if one abandons\footnote{See the
2nd reference in \cite{us}. This is related to the covariant formulations
of the antibracket formalism \cite{covariant}.} the assumption
of flat measures for the fields $\phi^A$, and if one does not restrict
oneself to local transformations that leave the functional measure
invariant. Some steps in this direction were recently taken in ref.
\cite{us1}. One here exploits the reparametrization invariance encoded
in the path integral by performing field transformations $\phi^A =
g^A(\phi',a)$ depending on new fields
$a^i$. It is natural to assume that these transformations form a
group, or more precisely, a quasigroup \cite{Batalin}. The objects
\beq
u^A_i ~\equiv~ \left.\frac{\delta^r g^A}{\delta a^i}\right|_{a=0}
\eeq
are gauge generators of this group. They satisfy
\beq
\frac{\delta^r u^A_i}{\delta\phi^B}u^B_j - (-1)^{\epsilon_i\epsilon_j}
\frac{\delta^r u^A_j}{\delta\phi^B}u^B_i = - u^A_k U^k_{ij} ~,\label{u}
\eeq
where the $U^k_{ij}$ are structure ``coefficients'' of the group. They
are supernumbers with the property
\beq
U^k_{ij} = - (-1)^{\epsilon_i\epsilon_j}U^k_{ji} ~.
\eeq

\noi
In ref. \cite{us1}, specializing to compact supergroups for which
$(-1)^{\epsilon_i}U^i_{ij} = 0$, the following $\Delta$-operator was derived:
\beq
\Delta G \equiv (-1)^{\epsilon_i}\left[\frac{\delta^r}{\delta\phi^A}
\frac{\delta^r}{\delta\phi^*_i}G\right]u^A_i + \frac{1}{2}(-1)^{\epsilon_i+1}
\left[\frac{\delta^r}{\delta\phi^*_j}\frac{\delta^r}{\delta\phi^*_i}G\right]
\phi^*_k U^k_{ji} ~. \label{rank-one-delta}
\eeq
When the coefficients $U^k_{ij}$ are constant, this $\Delta$-operator
is nilpotent: $\Delta^2 = 0$. As noted by Koszul \cite{Koszul}, and
rediscovered by Witten \cite{Witten},
one can define an antibracket $(F,G)$ by the rule
\beq
\Delta(FG) = F(\Delta G) + (-1)^{\epsilon_G}(\Delta F)G
+ (-1)^{\epsilon_G}(F,G) ~. \label{abdef}
\eeq
Explicitly, for the case above, this leads to the following new
antibracket \cite{us1}:
\beq
(F,G) \equiv (-1)^{\epsilon_i(\epsilon_A+1)}\frac{\delta^r F}
{\delta\phi^*_i}u^A_i\frac{\delta^l G}{\delta\phi^A} - \frac{\delta^r F}
{\delta\phi^A}u^A_i\frac{\delta^l G}{\delta\phi^*_i} + \frac{\delta^r F}
{\delta\phi^*_i}\phi^*_k U^k_{ij}\frac{\delta^l G}{\delta\phi^*_j}
\eeq

\noi
This antibracket is statistics-changing,
$\epsilon((F,G)) = \epsilon(F) + \epsilon(G) + 1$, and has the following
properties:
\begin{eqnarray}
(F,G) &=& (-1)^{\epsilon_F\epsilon_G+\epsilon_F+\epsilon_G}(G,F)
\label{exchange} \\
(F,GH) &=& (F,G)H + (-1)^{\epsilon_G(\epsilon_F+1)}G(F,H) \cr
(FG,H) &=& F(G,H) + (-1)^{\epsilon_G(\epsilon_H+1)}(F,H)G \label{Leibniz}
\\
0 &=& (-1)^{(\epsilon_F+1)(\epsilon_H+1)}(F,(G,H))
+ {\mbox{\rm cyclic perm.}} ~.\label{Jacobi}
\end{eqnarray}
Furthermore,
\beq
\Delta (F,G) = (F,\Delta G) - (-1)^{\epsilon_G}(\Delta F,G) ~.\label{dfg}
\eeq

\noi
The $\Delta$ given in eq. (\ref{rank-one-delta}) is clearly
a non-Abelian generalization of the conventional $\Delta$-operator of the
Batalin-Vilkovisky formalism.

\noi
We shall now show how to extend this construction to the general
case of non-linear and open algebras. Recently, interest in
more complicated algebras such as strongly homotopy Lie algebras
\cite{Stasheff} has arisen in the context of string field theory
\cite{Zwiebach}.

\noi
Consider the quantized Hamiltonian BRST operator $\Omega$ for first-class
constraints with an arbitrary, possibly open,
gauge algebra \cite{BF}.\footnote{For
a comprehensive review of the classical Hamiltonian BRST formalism, see,
$e.g.$, ref. \cite{Henneaux}.} Apart from a set of phase space operators
$Q^i$ and $P_i$, introduce a ghost pair $\eta^i,
{\cal P}_i$. They have Grassmann parities $\epsilon(\eta^i) =
\epsilon({\cal P}_i) = \epsilon(Q^i) + 1 \equiv \epsilon_i + 1$, and are
canonically conjugate with respect to the usual graded commutator:
\beq
[\eta^i,{\cal P}_j] =  \eta^i{\cal P}_j - (-1)^{(\epsilon_i+1)(\epsilon_j+1)}
{\cal P}_j\eta^i = i\delta^i_j ~.\label{supercom}
\eeq
In addition $[\eta^i,\eta^j] = [{\cal P}_i,{\cal P}_j] = 0$.
The quantum mechanical BRST operator can then be written in the form of
a ${\cal P}\eta$ normal-ordered expansion in powers of the ${\cal P}$'s
\cite{BF}:
\beq
\Omega = G_i\eta^i + \sum_{n=1}^{\infty}{\cal P}_{i_{n}}\cdots
{\cal P}_{i_{1}} U^{i_{1}\cdots i_{n}} ~.\label{Omega}
\eeq
Here
\beq
U^{i_{1}\cdots i_{n}} = \frac{(-1)^{\epsilon^{i_{1}\cdots i_{n-1}}_{j_{1}
\cdots j_{n}}}}{(n+1)!}U^{i_{1}\cdots i_{n}}_{j_{1}
\cdots j_{n+1}}\eta^{j_{n+1}}\cdots\eta^{j_{1}} ~,
\eeq
and the sign factor is defined by:
\beq
\epsilon^{i_{1}\cdots i_{n-1}}_{j_{1}
\cdots j_{n}} = \sum_{k=1}^{n-1}\sum_{l=1}^{k}\epsilon_{i_{l}} +
\sum_{k=1}^n \sum_{l=1}^k \epsilon_{j_{l}} ~.
\eeq
The $U^{i_{1}\cdots i_{1}}_{j_{n}\cdots j_{n+1}}$'s are generalized
structure coefficients. For rank-1 theories the expansion ends with
the 2nd term, involving the usual Lie algebra structure coefficients
$U^k_{ij}$. The number of terms that must be included in the expansion
of eq. (\ref{Omega}) increases with the rank. By construction
$\Omega^2 = 0$.

\noi
The functions $G_i$ appearing in eq. (\ref{Omega}) are the constraints.
In the quantum case they satisfy the constraint algebra
\beq
[G_i,G_j] ~=~ iG_k U^k_{ij} ~.
\eeq
We choose these to be the ones associated with motion on the
supergroup manifold defined by the transformation $\phi^A =
g^A(\phi',a)$.

\noi
When considering representations of the (super) Heisenberg algebra
(\ref{supercom}), one normally chooses the operators to act to the
right. Thus, in the ghost coordinate representation we could take
\beq
{\cal P}_j = i (-1)^{\epsilon_j}\frac{\delta^l}{\delta\eta^j} ~,
\eeq
and similarly in the ghost momentum representation we could take
\beq
\eta^j = i \frac{\delta^l}{\delta {\cal P}_j} ~.
\eeq

\noi
On the other hand, the most convenient representation of the constraint
$G_j$ is \cite{Batalin}
\beq
\lG_j = - i \frac{\ldr}{\delta \phi^A}u^A_j ~,
\eeq
which involves a right-derivative {\em acting to the left}.
Using eq. (\ref{u}), $\lG_j$ is seen to satisfy
\beq
[\lG_i,\lG_j] = i \lG_k U^k_{ij} ~.
\eeq

\noi
Since we wish $\Omega$ of eq. (\ref{Omega}) to act in a definite way,
we choose representations of the (super) Heisenberg algebra (\ref{supercom})
that involve operators acting to the left as well. These are
\beq
\lP_j = i \frac{\ldr}{\delta\eta^j}
\eeq
in the ghost coordinate representation, and
\beq
\leta_j = i (-1)^{\epsilon_j}\frac{\ldr}{\delta{\cal P}_j}
\eeq
in the ghost momentum representation.
Inserting these operators into eq. (\ref{Omega}) will give the corresponding
BRST operator $\lOmega$ acting to the left. We now identify the ghost
momentum ${\cal P}_j$ with the Lagrangian
antighost (``antifield'') $\phi^*_j$.

\noi
As a special case, consider
the operator $\lOmega$ in the case of an ordinary rank-1 super Lie algebra
for which $(-1)^{\epsilon_i}U^i_{ij} = 0$.
In the ghost momentum representation $\lOmega$ takes the form
\beq
\lOmega = (-1)^{\epsilon_i}\frac{\ldr}{\delta\phi^A}u^A_i\frac{\ldr}
{\delta\phi^*_i}
- \frac{1}{2}(-1)^{\epsilon_j}\phi^*_k U^k_{ij}\frac{\ldr}{\delta
\phi^*_j}\frac{\ldr}{\delta\phi^*_i} ~.
\eeq
One notices that the $\lOmega$ of the above equation
coincides with our non-Abelian $\Delta$-operator of eq. (
\ref{rank-one-delta}). In detail:
\beq
\Delta F ~\equiv~ F\lOmega ~. \label{delta-omega}
\eeq

\noi
For a rank-0 algebra -- the Abelian case --  we get,
with the same identification,
\beq
\lOmega = (-1)^{\epsilon_A}\frac{\ldr}{\delta\phi^A}\frac{\ldr}{
\delta\phi^*_A} ~.
\eeq
The associated $\Delta$-operator, defined through eq. (\ref{delta-omega})
is seen to agree with the $\Delta$ of the conventional
Batalin-Vilkovisky formalism \cite{BV}.

\noi
We define the general $\Delta$-operator through the
identification (\ref{delta-omega}) and the complete expansion
\beq
\lOmega = (-1)^i\frac{\ldr}{\delta\phi^A}u^A_i\frac{\ldr}{\delta\phi^*_i}
+ \sum_{n=1}^{\infty}\phi^*_{i_{n}}\cdots
\phi^*_{i_{1}} \lU^{i_{1}\cdots i_{n}} ~.\label{leftOmega}
\eeq
Here
\beq
\lU^{i_{1}\cdots i_{n}} = \frac{(-1)^{\epsilon^{i_{1}\cdots i_{n-1}}_{j_{1}
\cdots j_{n}}}}{(n+1)!}(i)^{n+1}(-1)^{\epsilon_{j_{1}} + \cdots
+ \epsilon_{j_{n+1}}}U^{i_{1}\cdots i_{n}}_{j_{1}
\cdots j_{n+1}}\frac{\ldr}{\delta\phi^*_{j_{n+1}}}\cdots\frac{\ldr}{
\delta\phi^*_{j_{1}}} ~.
\eeq
By construction we then have $\Delta^2 = 0$.

\noi
It is quite remarkable that the above derivation, based on Hamiltonian
BRST theory in the operator language,
has a direct counterpart in the Lagrangian path integral.
The two simplest cases, that of rank-0 and rank-1 algebras have been
derived in detail in the Lagrangian formalism in refs. \cite{us,us1}.
It is intriguing that completely different manipulations
(integrating out the corresponding ghosts $c^i$, and
partial integrations inside the functional integral) in the Lagrangian
framework leads to these quantized Hamiltonian BRST operators. The rank-0
case, that of the conventional Batalin-Vilkovisky formalism, corresponds
to the gauge generators
\beq
\lG_A = -i \frac{\ldr}{\delta\phi^A} ~.
\eeq
These are generators
of translations: when the functional measure is flat,
the Schwinger-Dyson BRST symmetry is generated
by local translations. The non-Abelian generalizations correspond to
imposing different symmetries as BRST symmetries in the path integral
\cite{us1}.

\noi
These non-Abelian BRST operators $\lOmega$ can be Abelianized by canonical
transformations involving the ghosts \cite{BF1}, but the significance of
this in the present context is not clear. Since in general the number
of ``antifields'' $\phi^*_i$ will differ from that of the fields
$\phi^A$, it is obvious that $u^A_i$ in general will be non-invertible.
Even when the number of antifields matches that of fields, the
associated matrix $u^A_B$ may be non-invertible (``degenerate'').\footnote{
In the special case where $u^A_B$ is invertible, the transformation
$\phi^*_A \to \phi^*_B(u^{-1})^B_A$ makes the corresponding $\Delta$-operator
Abelian \cite{us1}, but we are not interested in that case here. See
also refs. \cite{BT,Nersessian}.}

\noi
Having the general $\Delta$-operator available,
the next step consists in extracting
the associated antibracket. By the definition (\ref{abdef}), this
antibracket measures the failure of $\Delta$ to be a derivation. When
$\Delta$ is a second-order operator, the antibracket so defined will
itself obey the derivation rule (\ref{Leibniz}). For higher-order
$\Delta$-operators this is no longer the case. The antibracket will
then in all generality only obey the much weaker relation
\beq
(F,GH) = (F,G)H - (-1)^{\epsilon_G}F(G,H) + (-1)^{\epsilon_G}(FG,H) ~.
\eeq
The relation (\ref{dfg}) also holds in all generality.
When the $\Delta$-operator is of order three or higher, the antibracket
defined by (\ref{abdef}) will not only fail to be a derivation, but
will also violate the Jacobi identity (\ref{Jacobi}).

\noi
For higher-order $\Delta$-operators one can, as explained by Koszul
\cite{Koszul}, use the failure of the antibracket to be a derivation to
define {\em higher antibrackets}. These are Grassmann-odd analogues of
Nambu brackets \cite{Nambu,Takhtajan}. The construction is most
conveniently done in an iterative procedure, starting with the
$\Delta$-operator itself \cite{Koszul,Akman}. To this end, introduce
objects $\Phi^n_{\Delta}$ which are defined as follows:\footnote{Note
that our definitions differ slightly from ref. \cite{Koszul,Akman}
due to our $\Delta$-operators being based on right-derivatives, while those
of ref. \cite{Koszul,Akman} are based on left-derivatives.}
\begin{eqnarray}
\Phi^1_{\Delta}(A) &=& (-1)^{\epsilon_A}\Delta A \cr
\Phi^2_{\Delta}(A,B) &=& \Phi^1_{\Delta}(AB) - \Phi^1_{\Delta}(A) B
- (-1)^{\epsilon_A} A\Phi^1_{\Delta}(B) \cr
\Phi^3_{\Delta}(A,B,C) &=& \Phi^2_{\Delta}(A,BC) - \Phi^2_{\Delta}(A,B)C
- (-1)^{\epsilon_B(\epsilon_A+1)}B\Phi^2_{\Delta}(A,C) \cr
\cdot && \cdot \cr \cdot && \cdot \cr \cdot && \cdot \cr
\Phi^{n+1}_{\Delta}(A_1,\ldots,A_{n+1}) &=& \Phi^n_{\Delta}(A_1,
\ldots,A_nA_{n+1}) - \Phi^n_{\Delta}(A_1,\ldots,A_n)A_{n+1}\cr &&
- (-1)^{\epsilon_{A_{n}}(\epsilon_{A_{1}} + \cdots +\epsilon_{A_{n-1}} +1)}
A_n\Phi^n_{\Delta}(A_1,\ldots,A_{n-1},A_{n+1}) ~.
\end{eqnarray}

\noi
The $\Phi^n_{\Delta}$'s define the higher antibrackets. For example,
the usual antibracket is given by
\beq
(A,B) \equiv (-1)^{\epsilon_A}\Phi^2_{\Delta}(A,B) ~.
\eeq
The iterative procedure clearly stops at the first bracket that acts like
a derivation. For example, the ``three-antibracket'' defined by
$\Phi^3_{\Delta}(A,B,C)$ directly measures the failure of $\Phi^2_{\Delta}$
to act like a derivation. But more importantly, it also
measures the failure of the usual antibracket to satisfy
the graded Jacobi identity:\footnote{We thank K. Bering for pointing out an
error in the original version of this manuscript.}
\begin{eqnarray}
\sum_{\mbox{\rm cycl.}}(-1)^{(\epsilon_A+1)(\epsilon_C+1)}(A,(B,C))
&=& (-1)^{\epsilon_A(\epsilon_C+1)+\epsilon_B+\epsilon_C}\Phi^1_{\Delta}(
\Phi^3_{\Delta}(A,B,C)) \cr
&&+ \sum_{\mbox{\rm cycl.}}(-1)^{\epsilon_A(\epsilon_C
+1)+\epsilon_B+\epsilon_C}\Phi^3_{\Delta}(\Phi^1_{\Delta}(A),B,C) ~,
\end{eqnarray}
and so on for the higher brackets.

\noi
When there is an infinite number of higher antibrackets, the associated
algebraic structure is analogous to
a strongly homotopy Lie algebra $L_{\infty}$.
The $L_1$ algebra is then given by the nilpotent $\Delta$-operator,
the $L_2$ algebra is given by $\Delta$ and the usual antibracket,
the $L_3$ algebra by these two and the additional ``three-antibracket'',
etc. The set of higher antibrackets
defined above seems natural in closed string field theory
\cite{Zwiebach}, the corresponding $\Delta$-operator being
given by the string field BRST operator $Q$.

\noi
Having constructed the $\Delta$-operator (and its associated hierarchy
of antibrackets), it is natural to consider a quantum Master Equation of
the form
\beq
\Delta \exp\left[\frac{i}{\hbar}S(\phi,\phi^*)\right] = 0 ~.
\eeq
Using the properties of the $\Phi^{n}$'s defined above, we can write
this Master Equation as a series in the higher antibrackets,
\beq
\sum_{k=1}^{\infty}\left(\frac{i}{\hbar}\right)^k
\frac{\Phi^{k}(S,S,\ldots,S)}{k!} ~=~ 0 ~,\label{me}
\eeq
where each of the higher antibrackets $\Phi^{k}(S,S,\ldots,S)$ has $k$
entries. The series terminates at a finite order if the associated
BRST operator terminates at a finite order. For example, in the Abelian
case of shift symmetry the general equation (\ref{me}) reduces to
$i\hbar\Delta S - \frac{1}{2}(S,S) = 0$, the Master Equation of the
conventional Batalin-Vilkovisky formalism.

\noi
A solution $S$ to the general Master Equation (\ref{me}) is invariant under
deformations
\beq
\delta S = \sum_{k=1}^{\infty}\left(\frac{i}{\hbar}\right)^{k-1}
\frac{\Phi^{k}(\epsilon,S,S,\ldots,S)}{(k-1)!} ~,\label{Ssym}
\eeq
where again each $\Phi^{k}$ has $k$ entries, and $\epsilon$ is Grassmann-odd.
One can view this as the possibility of adding a BRST variation
\beq
\sigma\epsilon = \sum_{k=1}^{\infty}\left(\frac{i}{\hbar}\right)^{k-1}
\frac{\Phi^{k}(\epsilon,S,S,\ldots,S)}{(k-1)!} \label{mesym}
\eeq
to the action. Here $\sigma$ is the appropriately generalized ``quantum
BRST operator''.\footnote{For finite order, a rearrangement in terms of
increasing rather than decreasing orders of $\hbar$ may be more convenient.}
In the case of the Abelian shift symmetry, the above $\sigma$-operator
becomes $\sigma \epsilon = \Delta\epsilon + (i/\hbar)(\epsilon,S)$, which
precisely equals ($(i\hbar)^{-1}$ times) the quantum BRST operator of the
conventional Batalin-Vilkovisky formalism.

\noi
We note that the general Master Equation (\ref{me}) and the BRST symmetry
(\ref{Ssym}) has the same relation to closed string field theory
\cite{Zwiebach,Kugo} that the conventional Batalin-Vilkovisky Master
Equation and BRST symmetry has to open string field theory \cite{Witten}.
The r\^{o}le of the action $S$ is then played by the string field $\Psi$,
and the Master Equation (\ref{me}) is the analogue of the closed string
field equations. The symmetry (\ref{Ssym}) is then the analogue
of the closed string field theory gauge transformations.

\noi
The present definition of higher antibrackets suggests the existence
of an analogous hierarchy of Grassmann-even brackets based on a
supermanifold and a non-Abelian open algebra --  a natural generalization
of Possion-Lie brackets. It should also be interesting to
investigate the Poisson brackets and Nambu brackets
generated by the generalized antibrackets
and suitable vector fields $V$ anticommuting with the generalized
$\Delta$-operator (and in particular certain
Hamiltonian vector fields within the antibrackets),
as described in the case of the usual antibracket in ref. \cite{Nersessian1}.

\noi
So far our construction has been carried out in the ghost momentum
representation of the super Heisenberg algebra. But the definition of
an antibracket from the quantized Hamiltonian BRST operator can of
course be given in different representations on the extended
phase space spanned by $Q^i, P_j$ and $\eta^i,{\cal P}_j$. For example,
in the ghost coordinate representation, the BRST operator is a
first-order operator for Abelian and true Lie algebras (with just the
trivial ``one-bracket'' defined by it), but it becomes a higher order
operator suitable for defining higher antibrackets for general open
algebras. It is certainly a
challenge to find the r\^{o}le played by the associated antibracket
structure -- in particular in the ghost momentum representation --
in the Hamiltonian language.

\noi
We have restricted ourselves to field transformations
$\phi^A = g^A(\phi',a)$ that do not involve the ghosts $c^i$ or antighosts
$\phi^*_i$. Enlarging the transformations in this way should lead to a
fully covariant formulation of these non-Abelian antibrackets and
$\Delta$-operators. It is interesting to speculate, conversely, on the
meaning of the corresponding ``covariant'' Hamiltonian BRST operators.
In fact, the
analyses of ref. \cite{covariant} point, together with the present
observations, towards some surprising analogies in the Hamiltonian
and Lagrangian formulations. We hope some of these aspects can become
clarified in the future.

\vspace{0.5cm}

\noindent
{\sc Acknowledgement:}~ P.H.D. would like to thank A. Nersessian for
discussions, and S.L. Lyakhovich for pointing out ref. \cite{BT}.
The work of J.A. is partially supported by Fondecyt 1950809 and a
collaboration CNRS-CONICYT.

\end{document}